\documentclass[11pt,letterpaper]{article}
\usepackage{graphicx}
\usepackage{mathptmx}
\usepackage{amssymb}
\usepackage{bm}
\usepackage{times}
\usepackage{graphics}
 at 5truept 
\input prepictex.tex
\input pictex.tex
\input postpictex.tex
\def\fnote#1#2{\begingroup\def\thefootnote{#1}\footnote{#2}\addtocounter
{footnote}{-1}\endgroup}

\begin{document}
\baselineskip=12pt

%$\quad$
%\vskip36truept

\begin{center}
\begin{large}
{\bf Initial state propagators}
\end{large}
\end{center}
\vskip12truept

\begin{center}
Hael Collins\fnote{$\dagger$}{Electronic address:  hcollins@andrew.cmu.edu}  
\vskip4truept
{\it Department of Physics}
\vskip1truept

{\it Carnegie Mellon University, Pittsburgh, Pennsylvania\/}
\vskip1truept
{\small (August 9, 2013)}
\end{center}
\vskip6truept

\begin{abstract}
\noindent 
It is possible to define a general initial state for a quantum field by introducing a contribution to the action defined at an initial-time boundary.  The propagator for this theory is composed of two parts, one associated with the free propagation of fields and another produced by the operators of this initial action.  The derivation of this propagator is shown for the case of a translationally and rotationally invariant initial state.  In addition to being able to treat more general states, these techniques can also be applied to effective field theories that start from an initial time.  The eigenstates of a theory with interacting heavy and light fields are different from the eigenstates of the theory in the limit where the interactions vanish.  Therefore, a product of states of the noninteracting heavy and light theories will usually contain excitations of the heavier state once the interactions are included.  Such excitations appear as nonlocal effects in the effective theory, which are suppressed by powers of the mass of the heavy field.  By appropriately choosing the initial action, these excitations can be excised from the state leaving just effects that would be produced by a local action of the lighter fields.
\end{abstract}
\vskip24truept

\noindent 
{\sc Quantum\/} field theory is typically used for systems in simple quantum states.  For describing scattering processes this is appropriate.  To an excellent approximation the particles that participate in or result from a scattering process can be treated as the free-particle states of a noninteracting theory when looking long before or long after a collision has occurred.  Propagation is always made in reference to vacuum states defined in a far past and a distant future.  Being able to use these states brings many boons.  Subtleties such as which vacuum to use---the ground state of the free or of the interacting theory---largely do not matter, diagrammatic calculations permit many short-cuts such as the amputation of external legs, and the propagator which is the basis of the perturbative description of processes is of the simplest possible form.

However, in other physical settings being limited to only the asymptotic vacuum and the free particle states is a bit too restrictive.  It might not be practical, or even theoretically sound, in a particular system to define a state in an infinitely distant past.  Avoiding general excited states also means missing out on the possibility of describing other interesting dynamics---for example, systems which are not in an equilibrium state or the behaviour of quantum fields in the very early universe.

Given the central role of scattering processes in the development of quantum field theory, it is not surprising that the treatment of more complicated quantum states, other than the thermal state, have received far less attention.  But the importance of quantum fields in cosmology has encouraged an interest in a broader understanding of fields in more general quantum states.  Recently, for example, the authors of [1] have shown how to implement the choice of a particular initial state through an initial action of the fields.  Nonlocal operators in this initial action are related to the correlated structures and excitations of the field that are present at a particular moment.  The propagation of fields in this state are then affected by the presence of the operators in the initial action.

Our purpose with this article is to make the connection between a choice of an initial state and the form of the propagator a little clearer.  This connection of course can be viewed from either side:  we could either start with a particular state and derive its propagator or ask how we should choose the state so as to arrive at a particular propagator.  Along the way we shall also see how the standard interaction picture of quantum field theory generalizes so that it includes the effects of an initial state.  We illustrate these techniques with two examples.  We first show how to select an ordinary thermal state through an initial action.  The second example is a little more involved and shows how the ideas of effective field theory can be implemented in a theory with an initial time by choosing an appropriate initial state.
\vskip24truept

\noindent {\bf\large I. The propagator for the free vacuum\/} 
\vskip9truept

\noindent
Consider a theory of a spinless field in a flat space-time,
$$
S = \int dt \int d^3\vec x\, \bigl\{ 
{\textstyle{1\over 2}} \partial_\mu\phi\partial^\mu\phi 
- {\textstyle{1\over 2}} m^2\phi^2 - U(\phi) 
\bigr\} . 
$$
$U(\phi)$ is a potential from which the quadratic term has been removed.  We have written $U(\phi)$ as though it only depended on the field to make it simpler to go between the Lagrangian and Hamiltonian, but this is done mainly for illustration.  Nothing hinders us from including derivative interactions as well.

In the interaction picture, the free Hamiltonian determines how the field evolves,
$$
H_0 = \int d^3\vec x\, \bigl\{ 
{\textstyle{1\over 2}} \dot\phi^2 
+ {\textstyle{1\over 2}} \vec\nabla\phi\cdot\vec\nabla\phi 
+ {\textstyle{1\over 2}} m^2 \phi^2 
\bigr\} . 
$$
The theory can be equivalently expressed in a momentum representation as 
$$
H_0 = \int {d^3\vec k\over (2\pi)^3}\, \Bigl\{ 
{1\over 2} \dot\phi_{\vec k}\dot\phi_{-\vec k} 
+ {1\over 2} \omega_k^2\phi_{\vec k}\phi_{-\vec k} 
\Bigr\} , 
$$
where $\omega_k$ is the energy of an excitation, $\omega_k = \sqrt{k^2+m^2}$, writing $k=|\!|\vec k|\!|$.  The rest of the action, which contains all of the self-interactions of the field, 
$$
H_I = \int d^3\vec x\,\, U(\phi) ,
$$
generates the evolution of states.  So, for example, the evolution of the free vacuum state $|0(t)\rangle$ from $t'$ to $t$,
$$
|0(t)\rangle = U_I(t,t')\, |0(t')\rangle ,
$$
is produced by the following time-ordered operator,  
$$
U_I(t,t') = T e^{-i\int_{t'}^t dt^{\prime\prime}\, H_I(t^{\prime\prime})} 
= T e^{-i\int_{t'}^t dt^{\prime\prime}\int d^3\vec x\,\, U(\phi) } ,
$$
which is determined by the interacting parts of the theory, $H_I$.

We see that the time-dependence of the expectation value of an operator ${\cal O}(t)$ occurs in two places: in the operator itself and in both appearances of the state,
$$
\langle 0(t)|{\cal O}(t)|0(t)\rangle 
= \langle 0(t')|U_I^\dagger(t,t'){\cal O}(t) U_I(t,t')\, |0(t')\rangle .
$$
To evaluate the right side of this expression we resort to a standard bit of notational trickery.  We introduce a $\phi^+(t,\vec x)$ field associated with the $U_I(t,t')$ operator and a $\phi^-(t,\vec x)$ field associated with the $U_I^\dagger(t,t')$ operator.  The time-evolution of an expectation value of an operator starting in its free vacuum is then given by  
$$
\langle 0(t)|{\cal O}(t)|0(t)\rangle 
= \bigl\langle 0\big| T\bigl( {\cal O}^+(t) e^{-i\int_{t_0}^\infty dt'\, [H_I^+(t')-H_I^-(t')]} \bigr)\, \big|0\bigr\rangle .
$$
Here, $|0\rangle=|0(t_0)\rangle$ is a shorthand for the initial vacuum state.

Before going further it is worth pausing for a moment to make a few remarks about this expression.  (1) $H_I^\pm(t)$ means $H_I[\phi^\pm(t,\vec x)]$---the same potential as above is used, except that now it is being written entirely in $\phi^+$ or $\phi^-$ fields as indicated by the superscripts.  (2) The time-ordering is extended to apply to the $\phi^+$ and $\phi^-$ fields in the following sense:  $\phi^-$ fields always occur after the $\phi^+$ fields, whatever the specific numerical values of their arguments might be.  This ordering puts the $\phi^-$ fields to the left, which is the correct place for them; that is where the $U_I^\dagger(t,t')$ (with which they are associated) was in the first place.  (3)  The $\phi^-$ fields are further ordered in the opposite of the usual sense,
$$
T\bigl( \phi^-(t,\vec x)\phi^-(t',\vec y)\bigr) 
= \cases{ \phi^-(t',\vec y)\phi^-(t,\vec x) &if $t>t'$\cr
\phi^-(t,\vec x)\phi^-(t',\vec y) &if $t<t'$\cr}.
$$
This ordering is inherited from the Hermitian conjugation in $U_I^\dagger(t,t') = U_I(t',t)$; in effect, time runs backwards within this operator.  (4) The Hermitian conjugation is also responsible for the relative minus signs between the two Hamiltonians in the exponent.  (5) This exponent is integrated from $t_0$ to $\infty$.  The former is only a change in notation, $t'\to t_0$; but $t_0$ is better suited to denote an initial time.  Setting the upper limit to be $\infty$, rather than $t$, is a convenient, though ultimately unnecessary, convention:  nothing beyond $t$ ever contributes to the evolution.  The theory dutifully obeys causality.  (6) ${\cal O}(t)$ is an operator containing the fields $\phi$ and its derivatives which can be evaluated at different points.  ${\cal O}^+(t)$ means that all of the $\phi$'s have been replaced with $\phi^+$'s.  This is completely arbitrary.  ${\cal O}(t)$ appeared between $U_I^\dagger(t,t')$ and $U_I(t,t')$ so it could also have been written with the $\phi$'s replaced by $\phi^-$'s without altering the expectation value.

The appearance of an initial time creates complications which are usually avoided when setting up a scattering problem.  Let us call the vacuum of the free theory at $t=t_0$ $|0(t_0)\rangle$, and the vacuum of the interacting theory $|\Omega(t_0)\rangle$.  Speaking generally, these are not equal, and it would be wrong to assume that 
$$
\langle 0(t)|{\cal O}(t)|0(t)\rangle = \langle\Omega(t)|{\cal O}(t)|\Omega(t)\rangle ; 
$$
nor are we free, when $t_0$ is fixed and finite, to take $t_0\to -\infty$, to recreate Gell-Mann and Low's technique for projecting away the effects of excitations above the vacuum.  So when speaking of the initial state, it is important to be careful and to know what is really being meant.

When the interactions are small, the evolution of an expectation value can be found perturbatively by expanding the exponential and taking the Wick contractions of the fields.  There are two versions of fields, $\phi^\pm(t,\vec x)$, and as a consequence the Wick contractions lead to four propagators,
$$
\langle 0|T\bigl( \phi^\pm(t,\vec x)\phi^\pm(t,\vec y)\bigr) |0\rangle 
= G^{\pm\pm}(t,\vec x;t',\vec y) 
= \int {\textstyle{d^3\vec k\over (2\pi)^3}}\, e^{i\vec k\cdot(\vec x-\vec y)} 
G_k^{\pm\pm}(t,t') .
$$
The time-ordering in this expression obeys the extended sense described before, {\it viz.\/}
\begin{eqnarray}
G_k^{++}(t,t') &\!\!\!\!\!\!=\!\!\!\!\!\!& 
\Theta(t-t')\, G^>(t,t') + \Theta(t'-t)\, G^<(t,t') 
\nonumber \\
G_k^{+-}(t,t') &\!\!\!\!\!\!=\!\!\!\!\!\!& G^<(t,t') 
\nonumber \\
G_k^{-+}(t,t') &\!\!\!\!\!\!=\!\!\!\!\!\!& G^>(t,t') 
\nonumber \\
G_k^{--}(t,t') &\!\!\!\!\!\!=\!\!\!\!\!\!& 
\Theta(t'-t)\, G^>(t,t') + \Theta(t-t')\, G^<(t,t') . 
\nonumber 
\end{eqnarray}
Here $G^>(t,t')$ and $G^<(t,t')$ are the Wightman functions
\begin{eqnarray}
\langle 0|\phi(t,\vec x)\phi(t',\vec y) |0\rangle &\!\!\!\!\!\!=\!\!\!\!\!\!&
\int {\textstyle{d^3\vec k\over (2\pi)^3}}\, e^{i\vec k\cdot(\vec x-\vec y)} 
G_k^>(t,t') 
\nonumber \\
\langle 0|\phi(t',\vec y)\phi(t,\vec x) |0\rangle &\!\!\!\!\!\!=\!\!\!\!\!\!&
\int {\textstyle{d^3\vec k\over (2\pi)^3}}\, e^{i\vec k\cdot(\vec x-\vec y)} 
G_k^<(t,t') .
\nonumber 
\end{eqnarray}
For the free, Lorentz-invariant vacuum, these Wightman functions assume a familiar form,
$$
G_k^>(t,t') = {1\over 2\omega_k} e^{-i\omega_k(t-t')}
\qquad\hbox{and}\qquad 
G_k^<(t,t') = {1\over 2\omega_k} e^{i\omega_k(t-t')} .
$$
The propagator $G^{++}(t,\vec x;t',\vec y)$, for example, is at once recognized for the standard Feynman propagator, 
$$
G^{++}(t,\vec x;t',\vec y) =  
\int {d^3\vec k\over (2\pi)^3}\, {1\over 2\omega_k} \bigl\{ 
\Theta(t-t')\,  e^{-i\omega_k(t-t')} e^{i\vec k\cdot(\vec x-\vec y)} 
+ \Theta(t'-t)\, e^{i\omega_k(t-t')} e^{-i\vec k\cdot(\vec x-\vec y)}
\bigr\} .
$$
This expression could be converted into its more overtly Lorentz invariant form by introducing the appropriate contour integrals over $dk_0$; but since the physical problem here is to follow an expectation value as it evolves over finite intervals, starting at a specific initial time, nothing is to be gained by writing an integral over $d^4k$.  We integrate over all of space, but not (necessarily) over all of time.

In the following, the calculations have been typically written in terms of $G_k^>(t,t')$ and $G_k^<(t,t')$ rather than in terms of the specific Lorentz-invariant expressions.  This allows the results to be generalized with the effort of a moment to other space-times---though to keep the results reasonably tractable, the background here is always assumed to be translationally and rotationally invariant in the spatial directions.
\vskip24truept

\noindent {\bf\large II. The propagator for a more general initial state\/} 
\vskip9truept

\noindent
With these preliminaries now out of the way, we next consider a scalar field which at the initial time is in a more general state than its vacuum.  It is still possible to work in an interaction picture, but the choice of a different state means that the propagator will also change.  It is intuitively convenient to regard the propagator as having two parts:  a (free) vacuum part, telling how information in the field propagates from one space-time point to another once we have settled on a meaning for positive and negative energy eigenstates, which is independent of the state, and a second part that tells how the information contained in the initial state itself propagates forward.

A general state is specified by a density matrix $\rho(t)$.  Still working, as ever, in the interaction picture, this state is completely determined at later times once it has been specified at an initial time, $\rho_0=\rho(t_0)$.  It evolves through the same operator $U_I(t,t')$ defined before,
$$
\rho(t) = U_I(t,t_0) \rho_0 U_I^\dagger(t,t_0) . 
$$
Assuming that we have not clumsily left anything out of our theory, so that no probability is leaking away, $\rho(t)$ will be a Hermitian matrix.  It is always possible to choose a basis in which $\rho(t)$ is diagonal; but since we already have two distinct bases to consider---one for the free particle theory and one for the fully interacting theory---and since we do not wish to be bound to any particular choice for the basis from the very beginning, we shall not bother to assume that $\rho(t)$ has been diagonalized.

The expectation value of an operator ${\cal O}(t)$ is the weighted sum---a trace---over all the possible excitations contained in the state $\rho(t)$,
$$
\langle {\cal O}(t)\rangle_\rho 
\equiv {\rm tr}\, \bigl[ {\cal O}(t)\rho(t) \bigr] 
= {\rm tr}\, \bigl[ U_I^\dagger(t,t_0) {\cal O}(t)U_I(t,t_0) \rho_0 \bigr] . 
$$
The way that we shall fix the initial state is by defining a boundary action $S_0$ at $t=t_0$.  This idea was introduced in [1].  A particular configuration of the fields is then weighted by a factor $e^{iS_0}$.  The sum over all possibilities must be equal to 1, since we have been assuming that our theory captures everything relevant, so the density matrix for the initial state is written as 
$$
\rho_0 = Z^{-1}e^{iS_0} , 
$$
where $Z={\rm tr}\,\, e^{iS_0}$.  Any constant terms in the action can be absorbed into the normalization and we leave out the possibility of sources or sinks in the initial state.  This leaves the simplest structures in $S_0$ as the correlations between fields at two different points.  Still confining ourselves to states that are translationally and rotationally invariant, and assuming that the density matrix is real\footnote{Under Hermitian conjugation $\phi^+$ fields are replaced by $\phi^-$ fields and {\it vice versa\/}.  Everything else is complex conjugated.}, $\rho_0^\dagger=\rho_0$, the most general structure that is compatible with these restrictions and that is quadratic in the field is\footnote{Quadratic in the field, not necessarily its derivatives.  More generally, $S_0$ could also include structures depending on $\dot\phi(t_0,\vec x)$, $\ddot\phi(t_0,\vec x)$, {\it etc.\/}}  
\begin{eqnarray}
S_0 &\!\!\!\!\!\!=\!\!\!\!\!\!& 
- {1\over 2} \int d^3\vec x\, d^3\vec y\, \Bigl\{  
\phi^+(t_0,\vec x) A(\vec x-\vec y) \phi^+(t_0,\vec y) 
- \phi^-(t_0,\vec x) A^*(\vec x-\vec y) \phi^-(t_0,\vec y) 
\nonumber \\
&&\qquad\qquad\quad\  
+\,\, \phi^+(t_0,\vec x) iB(\vec x-\vec y) \phi^-(t_0,\vec y) 
+ \phi^-(t_0,\vec x) iB(\vec x-\vec y) \phi^+(t_0,\vec y) 
\Bigr\} . 
\nonumber 
\end{eqnarray}
$A(\vec x-\vec y)$ is a complex function and $B(\vec x-\vec y)$ is a real one.  The fact that the states have correlated structures over perhaps arbitrarily large distances is perfectly fine.  We do not know how the state was prepared beforehand, and there is in principle an infinite amount of time prior to $t_0$ for a clever experimenter or the universe to do so.  All of the details of how the state is prepared are irrelevant for what comes after.  The behaviour after $t_0$ is entirely determined by our knowledge of the state at this initial moment.

There is no {\it a priori\/} reason to stop at quadratic order in $S_0$, though there may be practical reasons to hope that only the quadratic terms are `large.'  Quadratic terms produce linear equations of motion.  Linear equations are much easier---which is usually meant to say, possible---to solve.  We shall therefore always imagine that any higher order correlations amongst the fields are small.  Then the operators on the boundary can be separated into two parts, and the nonlinearity of the theory can be solved perturbatively by extending the interaction picture in the following sense to include $S_0$:
\vskip6truept 

(1) The `free' part of the theory is declared to be the set of quadratic terms in the action, limited to those with no more than two derivatives, together with all of the quadratic terms in the boundary action.  This free theory defines the propagator.  In effective theories, the quadratic action can also contain higher derivative terms.  These are the remnants of nonlocalities produced by the propagation of heavier fields that have been left out of the effective theory.  Such nonlocalities are expanded in powers of the four-momentum and truncated at an appropriate order, determined by what can be resolved experimentally.  The derivative terms, beyond the standard kinetic term, are then naturally small in an effective theory and can be conveniently grouped amongst the interactions even though they too are quadratic in the field.
\vskip6truept 

(2) All the rest of the terms in the action, and all of the operators in $S_0$ that are cubic or higher order in the field are treated as interactions.  For this theory to be solvable in an approximate sense, both the sizes of the interactions in the action and the sizes of the three-point and higher correlators in the initial state must be small enough for a perturbative treatment to work.
\vskip6truept 

Thus the thing to compute is the propagator for the initial state.  Once it has been found, the perturbative calculation of any expectation value follows along its familiar course, aside from this change in the form of the propagator.  To do so, we cast out everything not needed for this purpose.  This leaves just the free part of the action, $H_0$ once again,
$$
H_0(t) = \int d^3\vec x\, \Bigl\{ {\textstyle{1\over 2}} \dot\phi^2 
+ {\textstyle{1\over 2}} \vec\nabla\phi\cdot\vec\nabla\phi 
+ {\textstyle{1\over 2}} m^2\phi^2 \Bigr\} ,
$$
and the quadratic terms of the initial state, which can also be cast into a Hamiltonian form,
\begin{eqnarray}
H_I(t) &\!\!\!\!\!\!=\!\!\!\!\!\!& 
\delta(t-t_0)\int d^3\vec x\, d^3\vec y\, \Bigl\{  
\phi^+(t_0,\vec x) A(\vec x-\vec y) \phi^+(t_0,\vec y) 
- \phi^-(t_0,\vec x) A^*(\vec x-\vec y) \phi^-(t_0,\vec y) 
\nonumber \\
&&\qquad\qquad\qquad\quad
+\,\, \phi^+(t_0,\vec x) iB(\vec x-\vec y) \phi^-(t_0,\vec y) 
+ \phi^-(t_0,\vec x) iB(\vec x-\vec y) \phi^+(t_0,\vec y) 
\Bigr\} , 
\nonumber 
\end{eqnarray}
by introducing a trivial time-dependence through the $\delta$-function.\footnote{We are using the convention where $\int_{t_0}^t dt'\, \delta(t'-t_0)\, f(t') = {1\over 2} f(t_0)$, when the $\delta$-function vanishes on the boundary of the integral.}

The notation being used here is deliberately suggestive:  the calculation will be performed in a slightly different version of the interaction picture than the one that we have just outlined.  For the time being, $H_0$ will be the free part and $H_I$ will be the interacting part of the Hamiltonian.  After we have found the propagator that includes the two-point information in the initial state, both these parts can be put back together and regarded as the free part of a more complicated theory, $H_0'=H_0+H_I$.  

The calculation could be performed by treating this $H_0'$ as the `free' Hamiltonian from the beginning, but this approach would leave a degree of freedom undetermined.  We are still obliged to say which are the positive energy states of the theory.  One advantage of building up the propagator from the vacuum propagator, is that we thereby define the sense of what is a positive energy excitation from the start.  If we feel that we have lost the freedom for making another choice of the energy states, it can be reintroduced by transforming the modes of the vacuum propagator.  Being careful and performing the calculation in a suitably general way, it is very simple to make such a transformation.  

With the logic of the calculation out of the way, we are ready to find the propagators appropriate for the initial states,
$$
\bigl\langle 0 \big| T\bigl( \phi^\pm(t,\vec x)\phi^\pm(t',\vec y) 
e^{-i\int_{t_0}^\infty dt'\, H_I(t')} \big| 0\bigr\rangle 
= \bigl\langle 0 \big| T\bigl( \phi^\pm(t,\vec x)\phi^\pm(t',\vec y) 
e^{iS_0} \big| 0\bigr\rangle . 
$$
The time integral over the interaction Hamiltonian only contributes at the initial boundary because of the $\delta$-function in $H_I(t)$.

The price of doing the calculation in this picture is that of course we must take the sum of an infinite series of corrections.  This turns out to be quite easy to do.  The idea is similar to one that is sometimes met as an elementary exercise in ordinary scattering theory.  Starting with an ordinary free, massive, spin zero particle, 
$$
\int d^4x\, \bigl\{ {\textstyle{1\over 2}}\partial_\mu\phi\partial^\mu\phi 
- {\textstyle{1\over 2}}m^2\phi^2 \bigr\} , 
$$
the massive propagator can be built from the massless one by treating the kinetic term as the free part of the theory and the mass as the interaction, and then summing an infinite series of corrections.  The propagator is the same as would have resulted from including the mass term in the free action from the start, as is more ordinarily done.  The calculation with an initial state is a bit more complicated for two reasons.  (1) There are seemingly four propagators to derive, not just one, and (2) there are more quadratic interactions to insert:  $\phi^+A\phi^+$, $\phi^-A^*\phi^-$, and $i(\phi^+B\phi^-+\phi^-B\phi^+)$.

The first apparent complication turns out not to be a problem at all.  The propagator for the initial state has a general form 
$$
G^{\pm\pm}_k(t,t') + \Delta G_k(t,t') .  
$$
The first term represents the vacuum propagator again while the latter is a common term shared by all of the possible choices of $\pm$ for the two fields.  Why the propagator has such a structure is easily understood.  The full propagator is found by summing an infinite series of connected diagrams: 
$$
\beginpicture
\setcoordinatesystem units <1truein,1truein>
\setplotarea x from -2.65 to 2.25, y from -0.15 to 0.12
\plot -1.5 0  -1.0 0 /
\plot -0.5 0   0.25 0 /
\plot  0.75 0  1.75 0 /
\put {{\footnotesize $\langle\phi^\pm(t,\vec x)\phi^\pm(t',\vec y)\rangle_{\rho_0} = $}} [r] at -1.62 0
\put {$+$} [c] at -0.75 0
\put {$+$} [c] at  0.50 0
\put {$+\ \ \cdots $} [l] at 1.9 0
\put {{\scriptsize $(t,\vec x)$}}  [c] at -1.50 -0.1
\put {{\scriptsize $(t',\vec y)$}} [c] at -1.00 -0.1
\put {{\scriptsize $(t,\vec x)$}}  [c] at -0.50 -0.1
\put {{\scriptsize $(t',\vec y)$}} [c] at  0.25 -0.1
\put {{\scriptsize $(t,\vec x)$}}  [c] at  0.75 -0.1
\put {{\scriptsize $(t',\vec y)$}} [c] at  1.75 -0.1
\put {$\bullet$} [c] at -0.125  0.00
\put {$\bullet$} [c] at 1.0625  0.00
\put {$\bullet$} [c] at 1.4375  0.00
\put {{\scriptsize $A,A^*\!\!\!,B$}} [c] at -0.125  0.07
\put {{\scriptsize $A,A^*\!\!\!,B$}} [c] at 1.0625  0.07
\put {{\scriptsize $A,A^*\!\!\!,B$}} [c] at 1.4375  0.07
\put {{\scriptsize $\pm$}} [c] at -1.5  0.07
\put {{\scriptsize $\pm$}} [c] at -1.0  0.07
\put {{\scriptsize $\pm$}} [c] at -0.5  0.07
\put {{\scriptsize $\pm$}} [c] at  0.25 0.07
\put {{\scriptsize $\pm$}} [c] at  0.75  0.07
\put {{\scriptsize $\pm$}} [c] at  1.75 0.07
\endpicture
$$
The first graph is the vacuum propagator.  In all the rest of the graphs, if the diagram is to be connected, the fields at $x=(t,\vec x)$ and $y=(t',\vec y)$ must be contracted with fields on the initial surface.  Because $t_0$ is either the earliest possible time (in the $+$ sense) or the latest possible time (in the $-$ sense), the Wightman function for $t,t'>t_0$ is determined entirely by the $\pm$ index of the field on the initial surface, 
$$
G_k^{\pm +}(t,t_0) = G_k^>(t,t_0), 
\qquad 
G_k^{\pm -}(t,t_0) = G_k^<(t,t_0).
$$
This establishes that the sum of all the graphs that connect to the initial time---what we called $\Delta G_k(t,t')$---is the same for all the propagators, irrespective of the $\pm$ indices.

There is a second simplification for the internal propagators within a graph.  An internal propagator connects points confined to the initial surface.  Whenever the times in a propagator are the same, all four forms are the same, $\tilde G_k \equiv G_k^{\pm\pm}(t_0,t_0)$.  Thus, only four time-dependent structures can appear,
$$
G_k^>(t,t_0)G_k^>(t',t_0),\quad
G_k^<(t,t_0)G_k^<(t',t_0),\quad
G_k^>(t,t_0)G_k^<(t',t_0),\quad\hbox{and}\quad
G_k^<(t,t_0)G_k^>(t',t_0) . 
$$
Which one occurs depends on which of the internal fields, those at $t_0$, are being contracted with the external fields, those at $t$ or $t'$.

Summing the series of connected graphs with insertions of the initial state operators produces the following correction to the propagator,
\begin{eqnarray}
\Delta G_k(t,t')&\!\!\!\!\!\!=\!\!\!\!\!\!&
{1\over 1 + i\tilde G_k[A_k-A_k^*+2iB_k]} \Bigl\{ 
- i \bigl[ G_k^>(t,t_0)G_k^>(t',t_0)A_k - G_k^<(t,t_0)G_k^<(t',t_0)A_k^* \bigr] 
\nonumber \\
&&\qquad\quad
-\,\, 
\bigl[ G_k^>(t,t_0)G_k^>(t',t_0) + G_k^<(t,t_0)G_k^<(t',t_0) \bigr] \tilde G_k 
\bigl[ |A_k|^2 - B_k^2 \bigr] 
\nonumber \\
&&\qquad\quad
+\,\, \bigl[ G_k^>(t,t_0)G_k^<(t',t_0) + G_k^<(t,t_0)G_k^>(t',t_0) \bigr]  
\bigl[ B_k + \tilde G_k \bigl[ |A_k|^2 - B_k^2 \bigr]\bigr] 
\Bigr\} . 
\nonumber 
\end{eqnarray}
This form can be used for any system that is translationally and rotationally invariant in the spatial directions.

In a Lorentz-invariant space-time, the Wightman functions assume the forms given earlier and $\tilde G_k = (2\omega_k)^{-1}$.  Expanding the complex function $A_k$ in terms of two real functions, $A_k=\alpha_k+i\beta_k$, the structure of the initial state affects the subsequent theory through the following addition to the propagator, 
\begin{eqnarray}
\Delta G_k(t,t')&\!\!\!\!\!\!=\!\!\!\!\!\!&
{1\over 4\omega_k^2} {1\over \omega_k-\beta_k-B_k} \Bigl\{ 
- 2\omega_k\alpha_k \sin\bigl[ \omega_k(t+t'-2t_0)\bigr] 
\nonumber \\
&&\qquad\qquad\qquad\quad
+ \bigl[ 2\omega_k\beta_k - \alpha_k^2 - \beta_k^2 + B_k^2 \bigr] 
\cos\bigl[ \omega_k(t+t'-2t_0)\bigr] 
\nonumber \\
&&\qquad\qquad\qquad\quad
+ \bigl[ 2\omega_k B_k + \alpha_k^2 + \beta_k^2 - B_k^2 \bigr] 
\cos\bigl[ \omega_k(t-t')\bigr] 
\Bigr\} . 
\nonumber 
\end{eqnarray}
\vskip24truept

\noindent {\bf\large III. Thermal states\/} 
\vskip9truept

\noindent
Before applying this method to something new, it is instructive to see first how it can be used to pick out a familiar state.  One of the simplest examples, after the free vacuum, is a thermal state.  An ensemble of spin zero fields at a temperature $T = 1/\beta$ is populated according to a Bose-Einstein distribution.  The number density $n_k$ of excitations with an energy $\omega_k$ follows a well known form,
$$
n_k = {1\over e^{\beta\omega_k} - 1} .
$$
The standard method for deriving the propagator for a thermal state is to apply a condition on the Wightman functions, which treats the temperature as an imaginary component of the time,\footnote{This is the KMS condition of Kubo, Martin, and Schwinger.} 
$$
G_k^>(t-i\beta,t') = G_k^<(t,t') . 
$$
When combined with the other boundary conditions on the Green's functions, the resulting thermal state propagator is 
$$
G^{\pm\pm}_k(t,t') + {n_k\over\omega_k} \cos\bigl[ \omega_k(t-t') \bigr] .
$$

Comparing it with the initial state propagator that we just derived, we see that this structure is arranged by choosing 
$$
\alpha_k = 0, \qquad
\beta_k = - 2\omega_k {n_k^2\over 2n_k+1}, \qquad
B_k = 2\omega_k {n_k(n_k+1)\over 2n_k+1} .
$$
This corresponds to choosing the following initial action, 
$$
S_0[\beta] = i \int {d^3\vec k\over (2\pi)^3}\, 
{\omega_k\over 2n_k+1} \biggl\{ 
n_k^2
\bigl( \phi^+_{\vec k}\phi^+_{-\vec k} + \phi^-_{\vec k}\phi^-_{-\vec k} \bigr)
- n_k(n_k+1) 
\bigl( \phi^+_{\vec k}\phi^-_{-\vec k} + \phi^-_{\vec k}\phi^+_{-\vec k} \bigr)
\biggr\} .
$$
In this case there is not an obvious reason to prefer fixing the thermal state thus rather than applying the KMS condition.  If anything, the latter is physically more intuitive; however the former can be applied to a far wider class of states.

This expression for the initial action differs slightly from how it is sometimes given elsewhere.  The initial action in terms of $\beta\omega_k$ rather than $n_k$ assumes the form
\begin{eqnarray}
S_0[\beta] &\!\!\!\!\!\!=\!\!\!\!\!\!& 
i \int {d^3\vec k\over (2\pi)^3}\, 
\biggl\{ {\omega_k\over 2\sinh\beta\omega_k} \Bigl[ 
\bigl[ \phi^+_{\vec k}\phi^+_{-\vec k} + \phi^-_{\vec k}\phi^-_{-\vec k} \bigr]
\cosh\beta\omega_k
-\bigl[ \phi^+_{\vec k}\phi^-_{-\vec k} + \phi^-_{\vec k}\phi^+_{-\vec k} \bigr]
\Bigr]
\nonumber \\
&&\qquad\qquad
-\,\, {1\over 2}\omega_k 
\bigl[ \phi^+_{\vec k}\phi^+_{-\vec k} + \phi^-_{\vec k}\phi^-_{-\vec k} \bigr]
\biggr\} .
\nonumber
\end{eqnarray}
This action corresponds to an implicitly normal-ordered theory.  The initial action for a theory that has not been normal-ordered would have been missing the final term,\footnote{This expression is a straightforward generalization of the quantum-mechanical density matrix in equation (10.44) of Feynman and Hibbs [2].}
$$
\hat S_0[\beta] = 
i \int {d^3\vec k\over (2\pi)^3}\, 
\biggl\{ {\omega_k\over 2\sinh\beta\omega_k} \Bigl[ 
\bigl[ \phi^+_{\vec k}\phi^+_{-\vec k} + \phi^-_{\vec k}\phi^-_{-\vec k} \bigr]
\cosh\beta\omega_k
-\bigl[ \phi^+_{\vec k}\phi^-_{-\vec k} + \phi^-_{\vec k}\phi^+_{-\vec k} \bigr]
\Bigr]
\biggr\} . 
$$
The final term in $S_0[\beta]$ was needed for the initial action to vanish in the zero temperature limit, $\beta\to\infty$ or $n_k\to 0$.  The initial action, $\hat S_0[\beta]$, in contrast contains excitations even in the $n_k\to 0$ limit,
$$
\hat S_0[\infty] = 
i \int {d^3\vec k\over (2\pi)^3}\, 
\Bigl\{ {\textstyle{1\over 2}} \omega_k 
\bigl[ \phi^+_{\vec k}\phi^+_{-\vec k} + \phi^-_{\vec k}\phi^-_{-\vec k} \bigr]
\Bigr\} . 
$$
Which is the correct initial action to use?  The presence of this term produces a correction to the vacuum propagator even in the limit where the temperature vanishes.  Therefore it is the normal-ordered $S_0[\beta]$ and not $\hat S_0[\beta]$ which is the proper initial action for selecting the standard thermal propagator.
\vskip24truept

\noindent {\bf\large IV. Simplifying the propagator\/} 
\vskip9truept

\noindent
Sometimes we have a particular initial physical state in mind for which we would like to determine the propagator.  But sometimes too, as the thermal state has illustrated, it is the reverse question that we wish instead to answer:  how should the initial state be chosen to produce a particular propagator?

For this purpose, the general form for the propagator associated with a rotationally and translationally invariant initial state can be parameterized by three momentum-dependent functions, $a_k$, $b_k$, and $c_k$, for the three corresponding independent time-dependent structures, 
$$
\Delta G_k(t,t') = 
{1\over\omega_k} \bigl\{ a_k \sin\bigl[ \omega_k(t+t'-2t_0)\bigr] 
+ b_k \cos\bigl[ \omega_k(t+t'-2t_0)\bigr] 
+ c_k \cos\bigl[ \omega_k(t-t')\bigr] 
\bigr\} . 
$$
We have defined these functions to be dimensionless by extracting a common factor of $1/\omega_k$.  A choice of the propagator---or equivalently, a choice for $a_k$, $b_k$, and $c_k$---in turn implies that a particular choice for $A_k=\alpha_k+i\beta_k$ and $B_k$ must have been made initially,
\begin{eqnarray}
\alpha_k &\!\!\!\!\!\!=\!\!\!\!\!\!&
- {2\omega_k\, a_k\over 1+2b_k+2c_k},  
\nonumber \\
\beta_k &\!\!\!\!\!\!=\!\!\!\!\!\!&
2\omega_k \biggl\{ 
b_k + {a_k^2\over 1+2b_k+2c_k} - {(b_k+c_k)^2\over 1+2b_k+2c_k} 
\biggr\} ,
\nonumber \\
B_k &\!\!\!\!\!\!=\!\!\!\!\!\!&
2\omega_k \biggl\{ 
c_k - {a_k^2\over 1+2b_k+2c_k} - {(b_k+c_k)^2\over 1+2b_k+2c_k} 
\biggr\} . 
\nonumber 
\end{eqnarray}
The relations between the two sets of real functions $\{ \alpha_k, \beta_k, B_k\}$ and $\{ a_k, b_k, c_k\}$ are nonlinear; but since the propagator was derived by summing an infinite series, such behaviour is hardly surprising.

The initial state action associated with this propagator is thus
\begin{eqnarray}
S_0 &\!\!\!\!\!\!=\!\!\!\!\!\!&
- {1\over 2} \int {d^3\vec k\over (2\pi)^3}\,  
\biggl\{ 
- {2\omega_k\, a_k\over 1+2b_k+2c_k} 
\bigl( \phi^+_{\vec k}\phi^+_{-\vec k} - \phi^-_{\vec k}\phi^-_{-\vec k} \bigr)
\nonumber \\
&&\qquad\qquad\qquad 
+\, {i\omega_k\, (b_k+c_k)\over 1+2b_k+2c_k} 
\bigl( \phi^+_{\vec k} + \phi^-_{\vec k} \bigr)
\bigl( \phi^+_{-\vec k} + \phi^-_{-\vec k} \bigr)
\nonumber \\
&&\qquad\qquad\qquad 
+\, i\omega_k \biggl[ b_k-c_k + {2a_k^2\over 1+2b_k+2c_k} \biggr]
\bigl( \phi^+_{\vec k} - \phi^-_{\vec k} \bigr)
\bigl( \phi^+_{-\vec k} - \phi^-_{-\vec k} \bigr)
\biggr\} .
\nonumber 
\end{eqnarray}
The first term could be put into a form that more closely resembles the other two through the identity
$$
\phi^+_{\vec k}\phi^+_{-\vec k} - \phi^-_{\vec k}\phi^-_{-\vec k}
= \bigl( \phi^+_{\vec k} + \phi^-_{\vec k} \bigr)
\bigl( \phi^+_{-\vec k} - \phi^-_{-\vec k} \bigr) . 
$$
For the thermal state, $a_k=0$, $b_k=0$, and $c_k=n_k$.
\vskip24truept

\noindent {\bf\large V. Matching for effective initial states\/} 
\vskip9truept

\noindent
We are now ready for a more interesting application.  Effective field theories are usually set up in an asymptotic vacuum state.  The initial and final times are sent to $t_0\to -\infty$ and $t\to +\infty$ respectively.  Suppose that instead we wish to use an effective theory that starts in an initial state at a particular initial time.  Intuitively we expect that in addition to the usual matching between the `complete' and `effective' actions of a theory, we might also need somehow to choose a suitable boundary action.  This `boundary matching'---what it means and how it is accomplished---is best illustrated with a simple example.

Consider a theory composed of a heavy field $\chi$ of mass $M$ and a light field $\phi$ of mass $m$.  At energies well below the mass of the heavier of the fields, it is possible to leave the heavy field out altogether and to describe all of the relevant dynamics of the light field in terms of an effective theory derived from the original theory.  The operators included in the effective theory are found by the familiar game of `matching' and `running':  we match the operators in the effective theory with those generated in the effective action of the original theory by the `one-light-particle-irreducible' graphs.\footnote{This is exactly as it sounds:  these are the graphs that are not cut in two by cutting a light virtual line.  They do not exclude graphs that be cut in two if we cut a heavy virtual line.}  This matching is done at the energy scale $\mu=M$.  It completely fixes the coefficients of the operators in the effective version of the theory.  Using the renormalization group, the coefficients are then run down to the lower energies where we intend to use this theory.

This standard treatment is always done in an asymptotic vacuum state.  Starting an interacting theory abruptly in a state means that, through the mixing of the two fields, it can contain excitations of the heavy fields, when expressed with respect to the basis of the free theory.  But the low energy effective theory should not include such excitations.  We need to find a way to remove them.  This can be done by adding a boundary codicil to our standard matching procedure.

We shall show the first perturbative step of this boundary matching through the following example\footnote{Some of the details of this example are examined more fully in [3].  We have included here only those results of the calculations that are needed to figure out the leading `boundary matching condition'.}, a theory with a light $\phi$ and a heavy $\chi$ scalar field,
\begin{eqnarray}
S[\phi,\chi] &\!\!\!\!\!\!=\!\!\!\!\!\!& 
\int d^4x\, \bigl\{ 
{\textstyle{1\over 2}}\partial_\mu\phi\partial^\mu\phi
- {\textstyle{1\over 2}}m^2\phi^2 
- {\textstyle{1\over 24}}\lambda\phi^4 
\nonumber \\
&&\qquad\, 
+\,\, {\textstyle{1\over 2}}\partial_\mu\chi\partial^\mu\chi
- {\textstyle{1\over 2}}M^2\chi^2 
- {\textstyle{1\over 24}}\Lambda\chi^4 
- {\textstyle{1\over 2}}g\phi^2\chi^2 
\bigr\} .
\nonumber 
\end{eqnarray}
This action contains all of the renormalizable operators for the light and heavy fields, together with one interaction between them.  We assume that $m\ll M$ so that an effective theory of a light $\phi$ is sensible, and we have imposed an invariance under the transformations $\phi\to -\phi$ and $\chi\to -\chi$.

One of the luxuries in treating asymptotic states---the sorts of states used in scattering problems---is that we can use an LSZ reduction to amputate external propagator legs from graphs.  This trick does not necessarily work for the expectation value of an operator evolved for a finite interval.  We must resort to other measures.

One trick is to give the light field a classical expectation value, $\phi_0(t)$.  We can then learn about the renormalization properties of the theory by examining the one-point function of the fluctuations about $\phi_0(t)$.  Even with the more complicated time-dependence, the one-point function of the fluctuations stays relatively simple:  all the graphs have a single external leg for the propagator for the light field.  This leg can be amputated.  What remains after the amputation is an equation of motion for $\phi_0(t)$ which is supplemented by the quantum corrections generated by the interaction of $\phi$ with $\chi$ and with itself.

So far we have spoken of $\phi_0(t)$ as an expectation value, but we have not yet specified in which state $\phi(t,\vec x)$ is being evaluated.  Let us very na\" ively put both fields in their free vacuum states at $t_0$ and evaluate the one-point function for the fluctuations about $\phi_0(t)$, 
$$
\langle 0(t)|\phi(t,\vec x)|0(t)\rangle = \phi_0(t) .
$$
At leading order in $g$ and in heavy particle loops, this one-point function produces the following condition on $\phi_0(t)$, 
\begin{eqnarray}
\ddot\phi_0(t) + m^2\phi_0(t) + {1\over 6}\lambda\phi_0^3(t) 
+ {1\over 2}g\phi_0(t) \int {d^3\vec k\over (2\pi)^3}\, {1\over\omega_k}
\qquad\qquad &&
\nonumber \\
-\,\, {1\over 2}g^2 \phi_0(t) \int_{t_0}^t dt'\, \phi^2_0(t') 
\int {d^3\vec k\over (2\pi)^3}\, 
{\sin[2\omega_k(t-t')]\over\omega_k^2} 
+ \cdots &\!\!\!\!\!\!=\!\!\!\!\!\!& 0 . 
\nonumber 
\end{eqnarray}
Since our purpose is to find what the existence of the heavy particle requires of the effective theory, we have not bothered to write graphs at the same order that are composed entirely of light fields.\footnote{Because the same graphs appear in both the complete and effective versions of the theory:  they already match.}  The $\omega_k$ here comes from the Wightman functions for the heavy field, so $\omega_k=\sqrt{k^2+M^2}$ contains the mass of this field, $M$.  The last term still has an integral over the time.  Through an infinite series of partial integrations, we transform that term, and thus the equation for $\phi_0$, into 
\begin{eqnarray}
\ddot\phi_0(t) + \biggl[ m^2 + {1\over 2}g \int {d^3\vec k\over (2\pi)^3}\, {1\over\omega_k} \biggr] \phi_0(t) 
+ {1\over 6} \biggl[ \lambda + {3\over 2}g^2 \int {d^3\vec k\over (2\pi)^3}\, {1\over\omega_k^3} \biggr] \phi^3_0(t) 
\qquad &&
\nonumber \\
-\,\, {g^2\over 16\pi^2} \phi_0(t) \sum_{n=1}^\infty {(-1)^n\over 2^n} 
{(n-1)!\over (2n+1)!!} {1\over M^{2n}} {d^{2n}\phi_0^2\over dt^{2n}}
\qquad &&
\nonumber \\
+\,\, 2g^2 \phi_0(t) \sum_{n=0}^\infty (-1)^n 
{d^{2n}\phi^2_0\over dt^{2n}}\Bigr|_{t_0} \int {d^3\vec k\over (2\pi)^3}\, 
{\cos[2\omega_k(t-t_0)]\over (2\omega_k)^{2n+3}} 
\qquad &&
\nonumber \\
+\,\, 2g^2 \phi_0(t) \sum_{n=0}^\infty (-1)^n 
{d^{2n+1}\phi^2_0\over dt^{2n+1}}\Bigr|_{t_0} \int {d^3\vec k\over (2\pi)^3}\, 
{\sin[2\omega_k(t-t_0)]\over (2\omega_k)^{2n+4}} 
+ \cdots &\!\!\!\!\!\!=\!\!\!\!\!\!& 0 . 
\nonumber 
\end{eqnarray}

Let us explain the meanings of the parts of this equation:
\vskip6truept 

(1) The mass $m^2$ and the coupling constant $\lambda$ both receive divergent corrections due to the heavy loops.  Despite the difference in the physical setting, any good field theorist will recognize these as structures that occur in the $S$-matrix analysis of this same theory.  If we dimensionally regularize these loop corrections, they become
$$
\int {d^3\vec k\over (2\pi)^3}\, {1\over\omega_k} 
= - {M^2\over 8\pi^2} \biggl[ {1\over\epsilon} - \gamma + \ln 4\pi 
+ 1 - \ln {M^2\over\mu^2} \biggr] 
$$
and
$$
\int {d^3\vec k\over (2\pi)^3}\, {1\over\omega_k^3}
= {1\over 4\pi^2} \biggl[ {1\over\epsilon} - \gamma + \ln 4\pi 
- \ln {M^2\over\mu^2} \biggr] , 
$$
which makes this connection more obvious.  For later it is useful to point out now that the correction to the mass has the structure 
$$
g \phi_0(t) \int {d^3\vec k\over (2\pi)^3}\, G_k^>(t,t) .
$$
This form, where we have written the general Wightman function and not its vacuum form, generalizes the more readily to the loop correction in a nonvacuum initial state.
\vskip6truept 

(2) Any good effective field theorist will also recognize the first infinite series of corrections in the equation for $\phi_0(t)$.  These are the relics of the propagation of heavy fields, a nonlocal effect that appears here in the guise of an infinite series of local operators.  Though they are terms in an equation for $\phi_0(t)$ and not in the form of an effective action for the quantum field $\phi(t,\vec x)$, our effective field theorist is clever enough to see that they require a series of derivative operators of the form 
$$
{c_n\over M^{2n}}\phi^2 (\square^2)^n \phi^2 
$$
in the effective theory.  Since we have not given $\phi_0(t)$ any spatial dependence, only the time-derivatives appear in the $\phi_0$-equation.  In the effective theory, we truncate this at some finite order, appropriate for what can be resolved experimentally.
\vskip6truept 

(3) Our intention was to put the theory in the free vacua of the light and heavy fields.  The last two lines in the equation for $\phi_0(t)$ show that we have been perhaps a little too na\" ive in thinking that this avoids any excitations of the heavy field.  The eigenstates of the free and interacting theories are not the same.  The terms in these two series are each nonlocal in time.  For example, the leading term, in powers of $1/M$, in the limit where $M(t-t_0)\gg 1$ is approximately
$$
{g^2\over 4} \phi_0(t) \phi^2_0(t_0) 
\int {d^3\vec k\over (2\pi)^3}\, {\cos[2\omega_k(t-t_0)]\over\omega_k^3} 
= {g^2 \phi^2_0(t_0)\over 8\sqrt{e\pi^3}} \phi_0(t) 
{\cos\bigl[ 2M(t-t_0) + {3\pi\over 4} \bigr]\over [2M(t-t_0)]^{3/2}} + \cdots . 
$$
The meaning of these terms is that the initial state contains excitations of the heavy field.  If we wait patiently, they will annihilate into light fields leaving only the usual operators in the effective theory for $\phi$.  That is guaranteed by the decaying $(t-t_0)^{-3/2}$ power law.  However, we could extend our matching prescription so that we include whatever structures are needed in the initial state to remove these nonlocal remnants of the heavy field completely.
\vskip6truept 

We illustrate how this matching is done for the initial state at order $g$ and at one-loop order.  We need to remove excitations of the heavy field, so we add a quadratic action for $\chi(t_0,\vec x)$ at the boundary.  The general form of the one loop correction to the mass term then becomes 
$$
g \phi_0(t) \int {d^3\vec k\over (2\pi)^3}\, G_k^>(t,t) 
\to g \phi_0(t) \int {d^3\vec k\over (2\pi)^3}\, G_k^>(t,t) 
+ g \phi_0(t) \int {d^3\vec k\over (2\pi)^3}\, \Delta G_k(t,t) 
$$
where the new term is 
$$
\int {d^3\vec k\over (2\pi)^3}\, \Delta G_k(t,t) 
= \int {d^3\vec k\over (2\pi)^3}\, 
{1\over\omega_k} \bigl\{ a_k \sin\bigl[ 2\omega_k(t-t_0)\bigr] 
+ b_k \cos\bigl[ 2\omega_k(t-t_0)\bigr] 
+ c_k
\bigr\} . 
$$
This cancels the one-loop, order $g^2$, nonlocalities when $c_k=0$ and 
\begin{eqnarray}
a_k &\!\!\!\!\!\!=\!\!\!\!\!\!& 
- 2g \sum_{n=0}^\infty {(-1)^n\over (2\omega_k)^{2n+3} } 
{d^{2n+1}\phi_0^2\over dt^{2n+1}}\Bigr|_{t_0} ,
\nonumber \\
b_k &\!\!\!\!\!\!=\!\!\!\!\!\!& 
- 2g \sum_{n=0}^\infty {(-1)^n\over (2\omega_k)^{2n+2} } 
{d^{2n}\phi_0^2\over dt^{2n}}\Bigr|_{t_0} . 
\nonumber 
\end{eqnarray}
At very high energies, $\omega_k\approx k$ and the leading behaviour of these functions is 
$$
a_k = - {g\over 4k^3} \phi_0(t_0)\dot\phi_0(t_0) + \cdots 
\quad\hbox{and}\quad 
b_k = - {g\over 4k^2} \phi_0^2(t_0) + \cdots .
$$
The resulting contribution from the initial state to loops is in general much more convergent at short distances than the contribution from the standard vacuum propagator.
\vskip12truept

\noindent {\bf Leading behaviour\/} 
\vskip3truept

Setting $c_k=0$ in the general expression for the boundary action produces
\begin{eqnarray}
S_0 &\!\!\!\!\!\!=\!\!\!\!\!\!&
- {1\over 2} \int {d^3\vec k\over (2\pi)^3}\, {2\omega_k\over 1+2b_k}
\biggl\{ 
- a_k 
\bigl( \chi^+_{\vec k}\chi^+_{-\vec k} - \chi^-_{\vec k}\chi^-_{-\vec k} \bigr)
\nonumber \\
&&\qquad\qquad\qquad\qquad\quad 
+\, i\, \bigl[ b_k + a_k^2 + b_k^2 \bigr]
\bigl( \chi^+_{\vec k}\chi^+_{-\vec k} + \chi^-_{\vec k}\chi^-_{-\vec k} \bigr)
\nonumber \\
&&\qquad\qquad\qquad\qquad\quad 
-\, i \bigl[ a_k^2 + b_k^2 \bigr]
\bigl( \chi^+_{\vec k}\chi^-_{-\vec k} + \chi^-_{\vec k}\chi^+_{-\vec k} \bigr)
\biggr\} .
\nonumber 
\end{eqnarray}
Both $a_k$ and $b_k$ are infinite series in $\phi_0$, so the boundary action has a complicated, nonlinear dependence on the expectation value $\phi_0$.  But by working in a limit appropriate for the effective theory, we can put this action into a form where its physical meaning is made a little more intuitive.

The effective theory is used when $\phi_0$ and its derivatives all are roughly the size of a common mass scale, which we shall call $E$, which is tiny when compared with the mass of the heavy field, $M$.  In this regime, the sizes of the terms in $a_k$ and $b_k$ form a series of powers of the small ratio $E/\omega_k\le E/M$.  In particular, $a_k$ contains odd powers of this ratio, and $b_k$ is made up only of even powers.  The leading term is thus the $n=0$ term of $b_k$, 
$$
b_k = - {g\over 2\omega_k^2}\phi_0^2(t_0) + \cdots \ll 1 .
$$
If we expand $S_0$ to leading order in this small quantity, we have as the leading boundary operator, 
$$
S_0 = {i\over 2} \int {d^3\vec k\over (2\pi)^3}\, {g\over \omega_k}\phi_0^2(t_0)
\bigl\{ 
\chi^+_{\vec k}(t_0) \chi^+_{-\vec k}(t_0)
+ \chi^-_{\vec k}(t_0) \chi^-_{-\vec k}(t_0) + \cdots 
\bigr\} .
$$
We can return from our long sojourn in momentum space by transforming back to position space, putting this leading part into a form from which we can extract its asymptotic behaviour in various limits,
\begin{eqnarray}
S_0 &\!\!\!\!\!\!=\!\!\!\!\!\!& 
{ig\over 4\pi^2} \int d^3\vec x\, d^3\vec y\, 
{M\over |\vec x-\vec y|} K_1\bigl( M|\vec x-\vec y| \bigr) \phi_0^2(t_0)
\nonumber \\
&&\qquad\qquad\quad\times
\bigl\{ 
\chi^+(t_0,\vec x) \chi^+(t_0,\vec y)
+ \chi^-(t_0,\vec x) \chi^-(t_0,\vec y) + \cdots 
\bigr\} .
\nonumber 
\end{eqnarray}
We now make a few remarks about the physical meaning of this action.
\vskip6truept 

(1) {\it Long distances:\/}  At large separations between the points $\vec x$ and $\vec y$, the Bessel function $K_1$ has the asymptotic form $K_1(z)\sim \sqrt{{\pi\over 2}} {e^{-z}\over\sqrt{z}}$.  The action is thus exponentially suppressed for distances much larger than the Compton wavelength of the heavy field.  This is a reasonable thing for an effective theory:  we do not expect large modifications of the long-distance properties of the effective theory by the presence of the heavy theory.  This does not mean that the state in the effective theory cannot have any long distance structures at all, only that they should not be associated with the heavy field.  We could have long distance structures in the effective state of light field if they are already there in the higher energy theory.
\vskip6truept 

(2) {\it Short distances:\/}  Even at short distances, the action is dominated by the local part of the operator since
$$
K_1\bigl( M|\vec x-\vec y| \bigr) = {1\over M|\vec x-\vec y|} + \cdots 
$$
for $M|\vec x-\vec y|\ll 1$.  The nonlocality of the operator is not able to be resolved by the dynamics of the light field at low energies.
\vskip6truept 

(3) {\it Dimensional analysis:\/}  A part of the point of giving the light field a background value is to learn as much as possible without the need to calculate higher-order correlation functions of the quantum field $\phi(t,\vec x)$ directly.  Earlier, we inferred the existence of the operators $\phi^2(\square^2/M^2)^n\phi^2$ in the effective theory from the presence of the $M^{-2n}\phi_0^2(d^{2n}/dt^{2n})\phi_0^2$ terms in the equation for the background.  We could have found the same operators by evaluating the four-point function of $\phi(t,\vec x)$, but it would have required considerably more effort.  Similarly, what we have been treating as a quadratic action in the heavy field should really be seen as the leading term in a tower of boundary operators in both fields.  This term in particular is a quartic operator, $\phi_0^2\chi^2\to \phi^2\chi^2$.  From the perspective of the {\it three-dimensional\/} Euclidean {\it boundary\/} field theory, this is an irrelevant operator---just what we should expect would be needed to cancel the unwanted short-distance parts of the state.  It is not a local operator of the boundary theory, but then we did not expect it to be so in the first place.

Once we start to consider an operator of the form $\phi^2\chi^2$, we are no longer looking at a `simple' modification of the two-point structure of the initial state.  Then we should be using the standard vacuum propagator\footnote{assuming that we are choosing the free vacuum state for the light field.  Even in the effective theory we are still free to consider initial states that contain low energy excitations of the light field.} while the modifications to the initial state are included amongst the interactions rather than the free parts of the theory.  This is only tractable as long as these interactions can be treated perturbatively.  But this is exactly what is happening here---in fact it is doubly so.  There is a suppression from the (presumably) small coupling $g$ and there is a further suppression by $\phi/\omega_k$, which is small in the limit where the effective theory is applicable.  It was in fact for this purpose, to postpone analysing such complications for as long as possible, that we introduced the background expectation value for the light field $\phi(t,\vec x)$ in the first place.
\vskip24truept

\noindent {\bf\large VI. A few last remarks\/} 
\vskip9truept

\noindent
Our purpose here has been to treat the evolution of matrix elements in a quantum theory that starts from a more general initial state and at an arbitrary time.  There are many interesting questions that might be addressed through this formalism.
\vskip6truept 

(1) The presence of interactions in a quantum theory leads to a renormalization of its parameters---masses, couplings, {\it etc\/}.  These same interactions might similarly alter the structure of the initial state, so it seems reasonable that the initial action will require some renormalization as well.  How is this renormalization performed?  Can we formulate a set of systematic rules for doing so?
\vskip6truept 

(2) An initial time might be naturally imposed by a physical system; but seen more generally, there could be some arbitrariness in the exact choice of this time.  If we have a particular state in mind, the time-evolution itself generates a flow in the parameters that we choose---{\it e.g.\/}~$A_k$ and $B_k$---to select it.  We can thus think of a flow in time, $t_0\to t'_0$, equivalently as a flow in the space of initial actions, $S_0(t_0)\to S'_0(t'_0)$, such that we stay within the same state.  From the perspective of the expectation value of an operator, $\langle{\cal O}(t)\rangle_\rho$, at times $t>t_0,t'_0$ it should not matter which we choose.  Are there any interesting properties about the theory that can be learned from this independence?  And what happens in an expanding background, such as de Sitter space, where an evolution in time is related to a flow in energies?
\vskip6truept 

(3) We have, in the second example, illustrated how the idea of an effective theory can be applied when starting at a finite initial time.  By allowing ourselves to choose more complicated initial states, the standard matching procedure of effective field theories needs to be extended to include a matching prescription for the boundary actions too.  We have only shown how this matching is done for the leading one-loop term that is linear in the coupling.  What is the general prescription for the boundary matching?  We would like to arrange a very clear matching procedure paralleling what is usually done to determine the operators in an effective theory.\footnote{For example, see the procedure explained in section 3.1 of [4]}  What the boundary matching is telling us is how to write the state perturbatively in the interacting theory so that it contains no excitations of the heavy fields.
\vskip6truept 

(4) These techniques can be used whenever we need to describe a quantum field that starts in a state other than an asymptotically defined one---a condensed matter system in an excited or nonequilibrium state or quantum fields in the very early universe, for example.  For the latter, we can avoid defining the quantum fluctuations in a nearly de Sitter background in the asymptotically distant past by simply defining it instead at a finite time in the past.  We can then methodically treat the possible influences of any earlier stages, or the effects of other fields, by considering various correlated structures in the initial action.  By comparing with observations, these structures can be constrained or eliminated empirically.
\vskip24truept

\centerline{\bf Acknowledgements}
\vskip9truept

\noindent 
I am grateful to the members of the Physics Department of Carnegie Mellon University for their hospitality and to Rich Holman for valuable discussions.
\vskip24truept

\centerline{\bf References}
\vskip12truept

\renewcommand{\theenumi}{[\arabic{enumi}]}
\renewcommand{\labelenumi}{\theenumi}

\begin{enumerate}
\item N.~Agarwal, R.~Holman, A.~J.~Tolley and J.~Lin, ``Effective field theory and non-Gaussianity from general inflationary states,'' JHEP {\bf 1305}, 085 (2013) [hep-th/1212.\-1172].
\item R.~P.~Feynman, A.~R.~Hibbs, {\it Quantum Mechanics and Path Integrals\/},
McGraw-Hill and Company, New York, 1965.
\item H.~Collins, R.~Holman and A.~Ross, ``Effective field theory in time-dependent settings,'' JHEP {\bf 1302}, 108 (2013) [hep-th/1208.3255].
\item H.~Georgi, ``Effective field theory,'' Ann.\ Rev.\ Nucl.\ Part.\ Sci.\  {\bf 43}, 209 (1993).
\end{enumerate}

\end{document}